\documentclass[default,iicol]{sn-jnl}

\usepackage{graphicx}%
\usepackage{multirow}%
\usepackage{amsmath,amssymb,amsfonts}%
\usepackage{amsthm}%
\usepackage{mathrsfs}%
\usepackage[title]{appendix}%
\usepackage[usenames,dvipsnames]{xcolor}
\usepackage{textcomp}%
\usepackage{manyfoot}%
\usepackage{booktabs}%
\usepackage{algorithm}%
\usepackage{algorithmicx}%
\usepackage{algpseudocode}%
\usepackage{listings}%
\usepackage[backend=bibtex,natbib=true,style=nature]{biblatex}
\usepackage{todonotes}
\usepackage{cleveref}

\definecolor{cream}{RGB}{222,217,201}

\newcommand{\reals}{\mathbb{R}}
\renewcommand{\SS}{\mathbb{S}}
\newcommand{\EE}{\mathbb{E}}
\newcommand{\X}{\mathcal{X}}
\newcommand{\C}{\mathcal{C}}
\newcommand{\D}{\mathcal{D}}
\newcommand{\given}{\,|\,}
\newcommand{\HD}{H_{\D}}
\newcommand{\HDnew}{H_{\D \cup (x,y)}}
\newcommand{\HDnewk}{H_{\D \cup (x,y_k)}}

\theoremstyle{thmstyleone}%
\begin{document}

\title[Article Title]{Probabilistic Prediction of Material Stability: Integrating Convex Hulls into Active Learning}

\author*[1]{\fnm{Andrew} \sur{Novick}}\email{novick@mines.edu}
\equalcont{These authors contributed equally to this work.}

\author[2]{\fnm{Diana} \sur{Cai}}\email{dcai@flatironinstitute.org}
\equalcont{These authors contributed equally to this work.}

\author[3]{\fnm{Quan} \sur{Nguyen}}\email{quan@wustl.edu}

\author[3]{\fnm{Roman} \sur{Garnett}}\email{garnett@wustl.edu}

\author[4]{\fnm{Ryan} \sur{Adams}}\email{rpa@princeton.edu}

\author[1]{\fnm{Eric} \sur{Toberer}}\email{etoberer@mines.edu}

\affil*[1]{\orgdiv{Department of Physics}, \orgname{Colorado School of Mines}, \orgaddress{ \city{Golden}, \state{Colorado}, \postcode{80401}, \country{USA}}}

\affil[2]{\orgname{Center for Computational Mathematics, Flatiron Institute}, \orgaddress{\city{New York}, \state{New York}, \postcode{10010}, \country{USA}}}

\affil[3]{\orgdiv{Department of Computer Science and Engineering}, \orgname{Washington University in St.~Louis}, \orgaddress{St.~Louis, Missouri, \postcode{63130}, \country{USA}}}

\affil[4]{\orgdiv{Department of Computer Science}, \orgname{Princeton University}, \orgaddress{\city{Princeton}, \state{New Jersey}, \postcode{08540}, \country{USA}}}

\abstract{Active learning is a valuable tool for efficiently exploring complex spaces, finding a variety of uses in materials science.  
However, the determination of convex hulls for phase diagrams does not neatly fit into traditional active learning approaches due to their global nature. 
Specifically, the thermodynamic stability of a material 
is not simply a function of its own energy, but rather requires energetic information from all other competing compositions and phases.
Here we present Convex hull-aware Active Learning (CAL), a novel Bayesian algorithm that chooses experiments to minimize the uncertainty in the convex hull. CAL prioritizes compositions that are close to or on the hull, leaving significant uncertainty in other compositions that are quickly determined to be irrelevant to the convex hull. The convex hull can thus be predicted with significantly fewer observations than approaches that focus solely on energy. Intrinsic to this Bayesian approach is uncertainty quantification in both the convex hull and all subsequent predictions (e.g., stability and chemical potential). By providing increased search efficiency and uncertainty quantification, CAL can be readily incorporated into the emerging paradigm of uncertainty-based workflows for thermodynamic prediction. 
}
 
\maketitle

\section{Introduction}\label{sec1}
Understanding thermodynamic stability is foundational to chemical and materials design. Phase relations provide mechanistic insight and accelerate discovery in disparate areas such as drug solubility \cite{williams2013strategies,baghel2016polymeric}, polymer blend stability \cite{robeson2007polymer,paul2012polymer,zhou2023correlating}, and phase transitions in metallic alloys \cite{okamoto1990binary,chen2018review}.
To accelerate stability predictions, computational research often focuses on producing high-fidelity surrogate models \cite{batzner2023advancing,drautz2019atomic,van2002automating,kadkhodaei2021cluster,GAP,bartok2015g}. However, phase stability prediction remains a persistent challenge for complex systems without effective surrogate models; examples include high-entropy materials \cite{cantor2004microstructural,george2019high,oses2020high,hart2021machine,HEO}, liquids and glasses \cite{zhuang2021like,mahanta2023local,therrien2021metastable}, materials at high temperatures \cite{griesemer2023accelerating}, and highly correlated materials \cite{dagotto2005complexity,keimer2017physics,kagomepressure,EuOGd,meschke2021search}. In this work, we address the frontiers of phase stability prediction by constructing an active learning approach that directly learns about the convex hull.

Phase transitions often occur across length- and time-scales too large to be directly observed using simulations. Instead, thermodynamic potentials need to be evaluated across a vast space of competing compositions and phases. The outcome of this competition is encapsulated in the convex hull: a single mathematical object that wraps the energy surface and defines the set of stable phase-composition pairs. Convex hulls are often associated with predicting the stability of compounds without external fields at 0\,K \cite{curtarolo2012aflow,oses2018aflow,OQMD,MaterialsProject,bartel2022review,merchant2023scaling,megahull,pandey2021predicting,rinkePerovskite}, but they have also been used to calculate phase transitions induced by temperature \cite{griesemer2023accelerating}, pressure \cite{yin1982theory,lee2020origins,jaffe2000lda}, anisotropic stresses in thin films \cite{xue2016strain,xue2017theory}, magnetic fields \cite{kuz2004spin,gorbunov2012magnetic}, and applied voltages in battery materials \cite{van1998first,van2020rechargeable}. Indeed, the convex hull formalism can be used to predict stability under any set of thermodynamic conjugate variables \cite{alberty2001use,sun2021generalized}. Beyond phase diagrams, convex hulls have been recently leveraged in understanding chemical reaction networks and synthesis pathways \cite{wen2023chemical,mcdermott2023assessing,rom2024mechanistically,chen2023navigating}.

\begin{figure*}[t]
\begin{center}
    \includegraphics[width =.95 \linewidth]{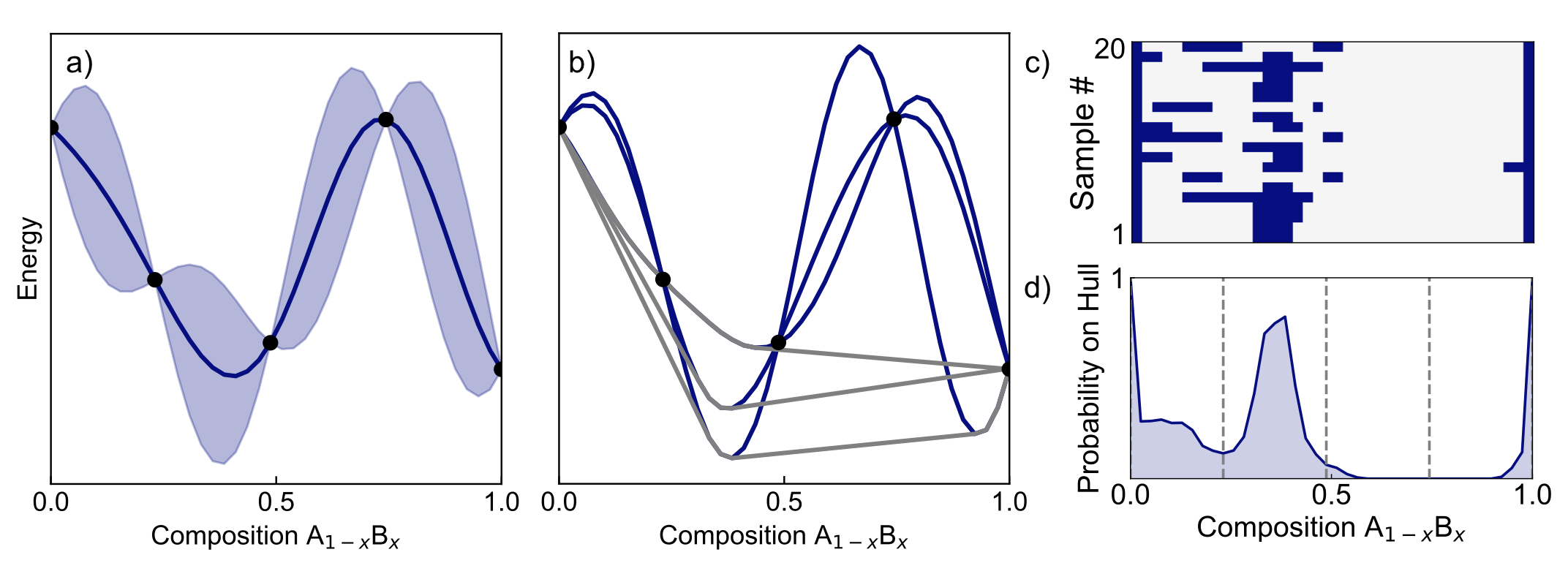}
\end{center}
\caption{\label{fig:Overview} (a) For a single phase, the search procedure begins by modeling the energy surface with a Gaussian process. The black points denote observed compositions, the blue curve represents the mean of the Gaussian process posterior, and the blue shaded region corresponds to two standard deviations from the mean. (b) Sampling from the Gaussian process posterior allows an ensemble of energy surfaces to be hypothesized. The convex hull (grey) is constructed for each energy surface; single-phase regions are where the energy surface touches the hull. (c) Each convex hull can be reduced to a composition vector with a binary classification of phase stability. Here, each row of the matrix corresponds to a separate sampled hull; blue denotes single phase compositions. (d) Interrogating this ensemble of hulls yields the probability of being on the hull. We note that observing the energy of compositions (dashed lines) does not necessarily give absolute information about their stability.}
\end{figure*}

The global nature of convex hulls implies that it is not obvious which composition-phase pairs will reside on the hull. For instance, it is possible for the exact value of the energy to be certain, while still being uncertain that the composition is on the hull. A brute force approach to predicting the convex hull would require calculating the energy for all competing phases and compositions. However, when the cost of individual energy evaluations is large, or the space of possible competing compositions is high-dimensional, exhaustively evaluating the energies is prohibitively expensive. Thus, there are two complimentary modes of acceleration: efficiently producing surrogate models that lower the cost of energy calculations and minimizing the number of energy evaluations necessary to define the convex hull. Both approaches can leverage active learning \cite{settles2009active}, since it is a natural method for selecting expensive data points that are expected to maximally increase the information about a function. 

To optimize the information gain about a surrogate energy function, active learning has been used to iteratively select first-principles calculations that minimize uncertainty in the surrogate model. Surrogate models like cluster expansion \cite{chen2024bayesian} and interatomic potentials \cite{gubaev2019accelerating} have been trained with active learning; they were then leveraged to conduct numerous energy evaluations for predicting the underlying convex hull. Active learning has also been biased to identify phase-composition pairs that are expected to be on or near the convex hull \cite{seko2020prediction,kuroda2023structural,vasylenko2024inferring}. While these approaches have been shown to be more efficient than random and grid-based search procedures, the active learning was only biased using  proxies that incorporate a local view of the hull rather than directly reasoning about the entire convex hull as a singular, global object. 

In this paper, we develop convex hull-aware active learning (CAL) to accelerate stability predictions. CAL distinguishes itself from more conventional Bayesian approaches
by reasoning directly about the entire convex hull. 
CAL uses separate Gaussian process regressions to model the energy surfaces of phases across the composition space. From the Gaussian processes, a posterior belief is produced over possible convex hulls. This induced posterior enables the algorithm to identify composition-phase pairs that are expected to minimize the uncertainty in the convex hull itself, not the constituent energy surfaces. 
By focusing exclusively on the convex hull, it is possible to make more effective decisions on what compositions to consider.

We start with illustrating the CAL algorithm in one dimension for clarity. The evolution of the convex hull distribution is seen with increasing observations, and  both stability predictions and chemical potentials are derived. From there, we explore complex ternary composition spaces with three competing phases. This allows us to  quantitatively demonstrate the efficiency of CAL against a baseline active learning procedure and explore analysis techniques for probabilistic hulls. 
 
\section{Approach}\label{sec2}

\begin{figure}[t!]
\begin{center}
\includegraphics[width =1 \linewidth]{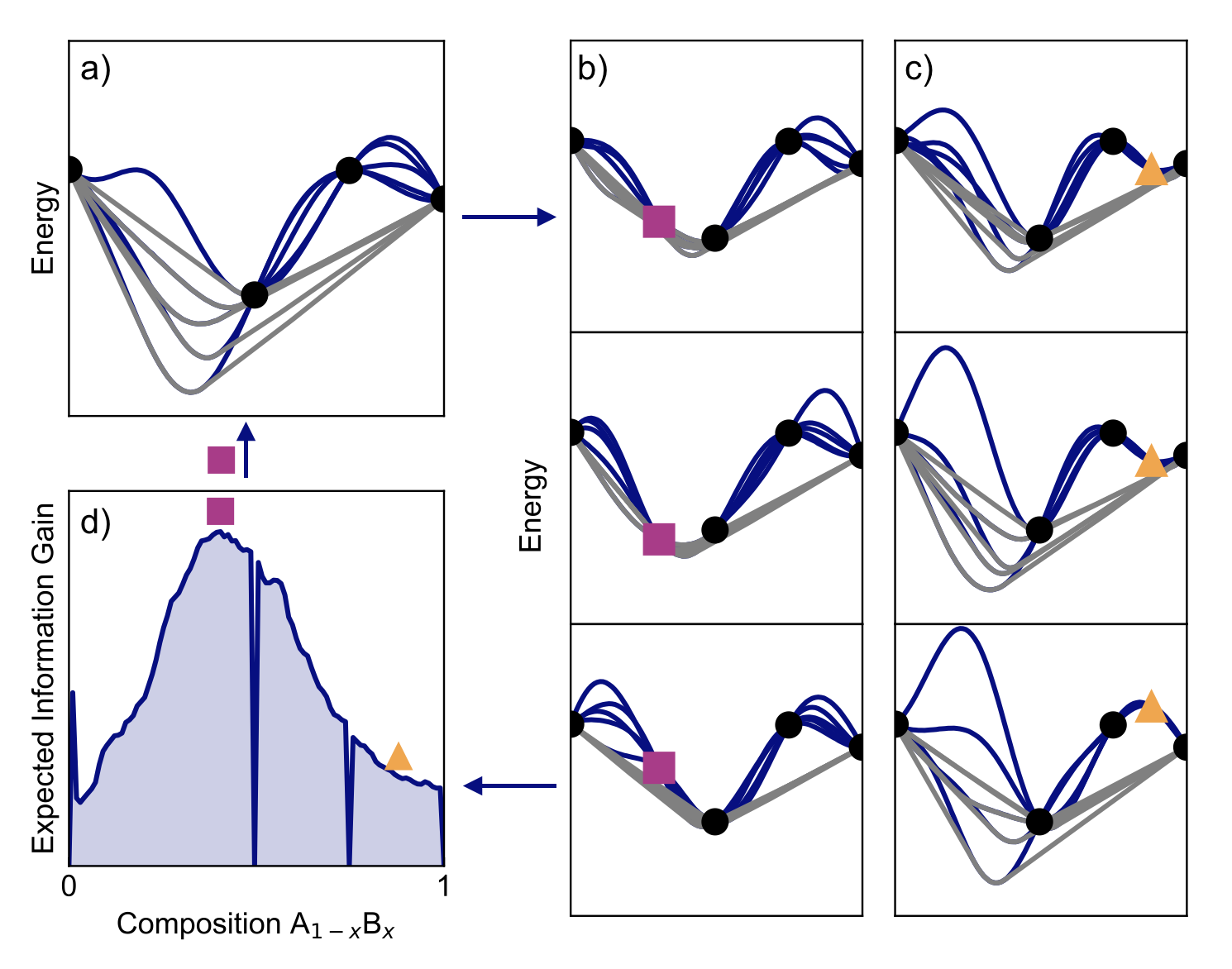}
\caption{\label{fig:Search Workflow} 
(a) Given some set of existing observations, energy surfaces are sampled from the trained GP and the corresponding hulls are calculated. (b)  To determine the expected information gain for a potential observation at composition $x'$, hypothetical energies that could result from such observations are predicted. These hypothetical energies are generated using the conditional distribution of the GP at $x=x'$. (c) For contrast, a  set of potential observations for a different $x$ composition are also highlighted. (d) This procedure is repeated to calculate the expected information gain across all compositions. The optimal composition $x \rightarrow{} x^*$ for subsequent observation is found by identifying the composition with the highest expected information gain. After conducting an observation at $x^*$, the process repeats until the uncertainty in the convex hull is sufficiently small.}
\end{center}
\end{figure}

The overall goal is to establish a methodology that approximates the convex hull with minimal observed data. We begin by establishing a probabilistic view of the hull (Fig.~\ref{fig:Overview}) and then present the policy for determining the next observation (Fig.~\ref{fig:Search Workflow}). 
We provide additional details on both the model and policy in the Methods section.

\paragraph{Probabilistic view of the hull}
In this and all subsequent examples, the energy surfaces are assumed to be continuous and differentiable across alloy compositions.
We also assume that there is a finite set of candidate compositions that represent a dense subset of the space.
In our first example, we begin with a single phase for which we have observed the energies of the parent compounds and three alloy compositions. These observations are denoted as~${\mathcal{D}=\{(x_n,y_n)\}^N_{n=1}}$, with~$x_n$ taking values in composition space and~$y_n$ being energies.

We model the energy surface with a
Gaussian process (GP),   which provides a prior on energy surfaces specified by a mean and covariance function \cite{GP,GPmatsci}.
Conditioning on the observations~$\mathcal{D}$ results in a posterior distribution over energy surfaces that is itself a Gaussian process (\cref{eq:gpmodel,eq:gpmodel2}).
Let~$F_{\mathcal{D}}$ be the random function associated with the posterior on energy surfaces; then~$H_{\mathcal{D}}=\C[F_{\mathcal{D}}]$ is the induced random (lower) convex hull, where~$\C$ is the convex hull operator.
The random function $H_{\mathcal{D}}$ is the object of primary interest  in this work.

As we are only considering a finite set of candidate compositions, it is possible to generate samples from this induced posterior by 1) drawing a sample from the multivariate Gaussian distribution resulting from the GP posterior, and 2) using a standard algorithm such as QuickHull \cite{qhull} for computing the lower convex hull of a set of points.
Fig.~\ref{fig:Overview}a shows a posterior distribution over the energy surface, $F_{\mathcal{D}}$, and Fig.~\ref{fig:Overview}b depicts three posterior samples and their associated convex hulls.

Our epistemic uncertainty about the true convex hull is captured by the random function~$H_{\mathcal{D}}$; the Shannon entropy~$\SS[H_{\mathcal{D}}]$ then quantifies our (lack of) knowledge about the convex hull.
By framing our problem as one of minimizing~~$\SS[H_{\mathcal{D}}]$, we can more rapidly gain information about the structure in which we are most interested.

In addition to the hull itself, various properties of interest can be derived from~$H_{\mathcal{D}}$, so we can reason about their posterior distributions as well.
For example, the (random) set
\begin{align*}
    \mathcal{S}_{\mathcal{D}} &:= \{ x: F_{\mathcal{D}}(x) = H_{\mathcal{D}}(x)\}
\end{align*}
contains the stable compositions as these are the compositions for which the minimum-energy phase is tight against the convex hull.

Fig.~\ref{fig:Overview}c shows 20 samples of stable sets after the 3 iterations in \ref{fig:Overview}b.
These binary classifications can be averaged to estimate the marginal probability that any given composition is on the hull, i.e., is stable (Figure \ref{fig:Overview}d).
Note that these marginal probabilities reveal an important way in which this problem is different from conventional Bayesian optimization and active learning tasks: the global nature of the convex hull means there is uncertainty about stability even for compositions in which the energy has been noiselessly observed.
In this example, the observed compositions are marked with dashed vertical lines in Fig.~\ref{fig:Overview}d and there is uncertainty about the stability in two of the three cases.

\paragraph{Refining the convex hull} 
With a probabilistic view of convex hulls in place, our goal in each iteration of the search is to identify the candidate observation~$x^*$, which is expected to minimize the Shannon entropy~$\SS[H_{\mathcal{D}}]$.
This objective can be viewed as a Bayesian experimental design procedure in which the policy is to greedily maximize the information gain (Fig.~\ref{fig:Search Workflow}).

Like many Bayesian optimization and search algorithms, the selection of $x^*$ requires approximating the expected information gain (EIG) across the space of possible designs, which in our case is the set of compositions \cite{mackay1992information,mackay1992evidence}.
The EIG is simply the difference between the Shannon entropy of the current state (reflected in the observed data, $\mathcal{D}$) and the expected Shannon entropy after making an observation at an unobserved composition~$x$.
Of course, the energy value~$y$ is unknown at this point and so the new set of observations~${\mathcal{D}\cup(x,y)}$ is considered in expectation:
\begin{align}
\label{eqn:eig}
    \text{EIG}(x\,;\,\mathcal{D}) &:=
    \SS[H_{\mathcal{D}}]
    -\mathbb{E}_{y}[\SS[H_{\mathcal{D}\cup(x,y)}]]\,.
\end{align}

Finally, the expected information gain is used within each iteration to select~$x^*$, the candidate composition to be evaluated:
\begin{align*}
x^* &= \arg\max_{x} \, \text{EIG}(x\,;\,\mathcal{D})\,.
\end{align*}

Fig.~\ref{fig:Search Workflow} illustrates how the EIG is evaluated in practice.
In Fig.~\ref{fig:Search Workflow}a, we start with a GP conditioned on some data,~$\mathcal{D}$.
Energy surfaces are sampled from the resulting posterior distribution, convex hulls are calculated, and the Shannon entropy of state~$\mathcal{D}$ is calculated, giving us the first term in equation~\ref{eqn:eig}.

For a given candidate composition~$x$, we sample from the conditional Gaussian process posterior at $x$ to obtain a set of~$K$ possible energy values, denoted~$y_k$. 
In other words, these $y_k$ values correspond to different energies for composition $x$ given the current uncertainty within our energy model.
For each of these ~$K$ samples, the entropy~$\SS[H_{\mathcal{D}\cup(x,y_k)}]$ is estimated in three steps.
1) The Gaussian process is conditioned on this ``fantasized'' pair of observations~$(x,y_k)$, and energy surfaces for all considered compositions are sampled from the resulting distribution.
2) For each of these sampled energy surfaces, a convex hull is computed.
3) The convex hull samples are used to estimate the Shannon entropy (\cref{eq:entropydef}), as detailed in the Methods. 
The expectation value of the Shannon entropy is then calculated by averaging the $K$ entropy estimates (\cref{eq:entropy}),
resulting in an estimate of
the second term in \cref{eqn:eig}, thereby completing our evaluation of the EIG. 

We continue to illustrate this algorithm in panels Fig.~\ref{fig:Search Workflow}b where three hypothetical energy values for composition $x$ lead to three different hull distributions. For contrast, a different composition is selected for Fig.~\ref{fig:Search Workflow}c, resulting in visibly greater variation in the hulls and thus a higher expected Shannon entropy.  In Fig.~\ref{fig:Search Workflow}d, the process is repeated across composition space to determine the composition with the maximum EIG (i.e., $x^*$).
(For panel b, the optimal value $x^*$ was intentionally selected to visually emphasize the impact that sampling at $x^*$ would have.)
Finally, an observation is made at $x^*$ to update $\mathcal{D}$ and the algorithm repeats to refine the convex hull. 
We reiterate that this approach seeks to minimize the Shannon entropy in the convex hull, not simply observe points that are on the hull. 
Here, observing the composition $x^*$ is advantageous because regardless of its energy, the resulting distribution in possible convex hulls narrows significantly. 

\begin{figure}
\begin{center}
\includegraphics[width =0.6 \linewidth]{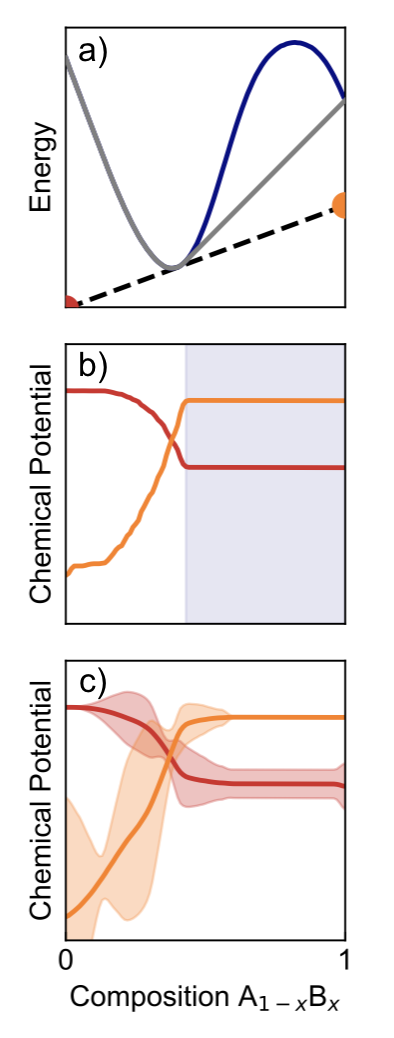}
\caption{\label{fig:Chemical Potential} 
(a) Given a sampled energy surface from the GP (blue), intensive properties can be obtained from the associated hull (grey); when considering $E(x)$, the tangent (black) to the hull yields the elemental chemical potentials upon intersection with $x=0$ and $x=1$, denoted by the red and orange points.  (b) For the single sampled hull, the chemical potentials are derived across the composition space. Within the two-phase region (shaded), the chemical potentials are constant. 
(c) From an ensemble of convex hull samples, the corresponding distribution in elemental chemical potentials are also represented as a distribution. The uncertainty in these potentials can be used to inform stopping criteria.  
}
\end{center}
\end{figure}

\begin{figure*}[t!]
\begin{center}
\includegraphics[width =0.67 \linewidth]{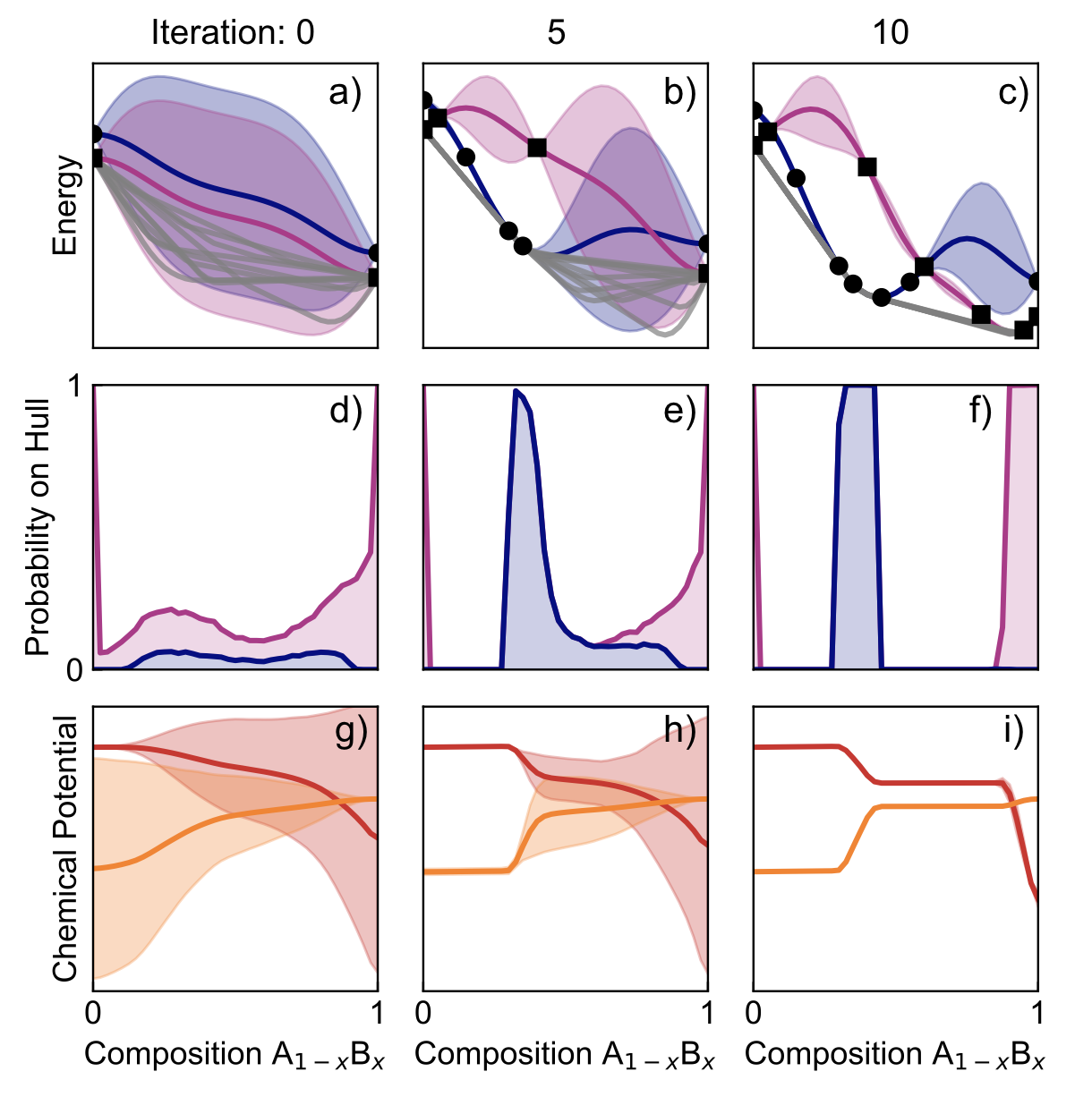}
\caption{\label{fig:1D Time Series} 
Chemical systems with multiple competing phases are represented with independent GPs; here,  two  phases (blue and purple) are considered in a binary space.
(a)  Having observed nothing but the endpoints, there is significant uncertainty across the composition space. Ten example convex hull samples are shown in grey, and they also vary widely. With (b) 5 and (c) 10 iterations, the distribution of hulls converges.  (d-f) The probability that a given phase is on the hull likewise converges with observation iterations. These are stacked plots such that the total probability for being on the hull is broken up into the individual phase contributions. (g-i) The elemental chemical potentials also converge after 10 iterations ($\mu_A$: red; $\mu_B$: orange).}
\end{center}
\end{figure*}
\paragraph{Application of the Convex Hull} 
Having sufficiently iterated to build an accurate hull, relevant thermodynamic intensive variables can be directly calculated.
For example, the elemental chemical potentials  can be determined by combining the tangent and energy value of the hull. Figure \ref{fig:Chemical Potential}a highlights that the elemental chemical potentials can be directly read off the $y$-intercepts of the composition boundaries (i.e., $x=0$ and $x=1$). Here, the energy surface is a single sample from a GP with an associated convex hull. Sweeping over the derivative of the convex hull changes the elemental chemical potentials, as shown in Figure \ref{fig:Chemical Potential}b. 
All compositions within the two-phase region (shaded in blue) are in thermodynamic equilibrium, and as such, the chemical potentials stay constant.  
Figure \ref{fig:Chemical Potential}c shows the mean chemical potential and affiliated uncertainty ($\pm2\sigma{}$) associated with a distribution of convex hulls. 

Elemental chemical potentials are critical in predicting defect concentrations, as defect creation involves exchanges with element and charge reservoirs. For example, in LiZnSb, the limited chemical potential window of Li renders the compound significantly Li-deficient even in the presence of secondary phases with excess Li (e.g. Li$_3$Sb) \cite{gorai2019LiZnSb}.  
Chemical potentials of charged species can also be leveraged to produce intercalation voltage curves in battery materials \cite{van2020rechargeable}, as was done for Li$_x$CoO$_2$ \cite{van1998first}.
Lastly, pressure is an intensive variable that can be determined from the convex hull of an energy surface that is a function volume \cite{yin1982theory,lee2020origins,jaffe2000lda}. For example, the impact of volumetric confinement on the freezing point of water can be readily determined from the hull \cite{powell2020freezing}.

\paragraph{Multiple phases}
CAL can be naturally expanded to search across multiple competing phases. In such cases, the $n$ phases are modeled with $n$ independent GPs.
By adopting separate GPs, we make no assumptions concerning correlations between the energy surfaces of different phases. For further efficiency, the set of $n$ phases could be described with a joint GP, as mentioned in the Discussion. 
To construct the corresponding convex hull distribution, each GP is sampled $s$ times, resulting in $s^n$ permutations of $n$ energy surfaces. 
For a given permutation, the $n$ energy surfaces, corresponding to the $n$ phases, can once again be wrapped with a single convex hull. From the convex hull we can predict the probability that a given phase-composition pair is on the hull, as will be shown in Fig.~\ref{fig:1D Time Series}.
The search process extends gracefully to multiple phases; the expected information gain is evaluated for each phase-composition pair.

\paragraph{Case Example I: 1D, 2 Phases}  To see this methodology applied to an iterative loop, we consider the case of a 1-dimensional binary composition space with two competing phases. 
Figure \ref{fig:1D Time Series}a shows how the initial energy surfaces are ambiguous and this uncertainty propagates to the convex hull. The probability of any composition being on the hull is then derived from the convex hull distribution.  
In  Fig.~\ref{fig:1D Time Series}b and c, increasing observations leads to a tightening of the energy and convex hull distributions. However, CAL leaves significant ambiguity in the energy surfaces when they are well above the hull.  
The probability of a given phase being on the hull is shown across Fig.~\ref{fig:1D Time Series}d-f; these curves quantify the evolving uncertainty in the stability predictions. A similar evolution is seen in the elemental chemical potentials (Fig.~\ref{fig:1D Time Series}g-i).

As previously mentioned, CAL acquires observations that minimize the uncertainty in the convex hull distribution. The behavior of the algorithm can be characterized by two steps. In the first few iterations when there is large uncertainty, Figure \ref{fig:1D Time Series}b shows that the algorithm tends to explore the energy surface, producing a coarse estimate for the convex hull. As the estimate of the convex hull develops, the algorithm focuses its next iterations increasingly on regions that are purportedly on the hull or close to it. These subtle refinements to the convex hull distribution are reflected in Figure \ref{fig:1D Time Series}c, where the convex hull samples converge.  

\paragraph{Quantitative performance assessment}   
Hulls are intriguing objects as they involve both classification and quantitative prediction. In part, we seek to classify if a given composition is on the hull. 
Knowing about the energies and slopes of the hull are also important for deriving intensive variables and quantifying the energy above the hull for an unstable composition. 
For this reason, we use three metrics in order to assess these dual aims: mean absolute error (MAE) for the hull energy, true positive rate (TPR), and false positive (FPR). Here, TPR refers to the percentage of stable compositions that are correctly identified as being on the hull, while FPR is the percentage of unstable compositions that are incorrectly identified as being on the hull. Mathematical definitions for these metrics can be found in the Methods. 
 
In low dimensions, producing an accurate hull can be achieved via brute force. However, the necessity for efficient hull construction emerges in spaces that involve multiple competing phases and large composition spaces. To test the efficiency of CAL in such a space, we pit it against a challenging opponent: a baseline algorithm (BASE) that still models the energy surfaces using a Gaussian process. However, BASE seeks to minimize the uncertainty in the energy surfaces and has no knowledge of convex hulls. See the Methods for further information about the BASE policy.

\begin{figure}
\begin{center}
\includegraphics[width =0.6 \linewidth]{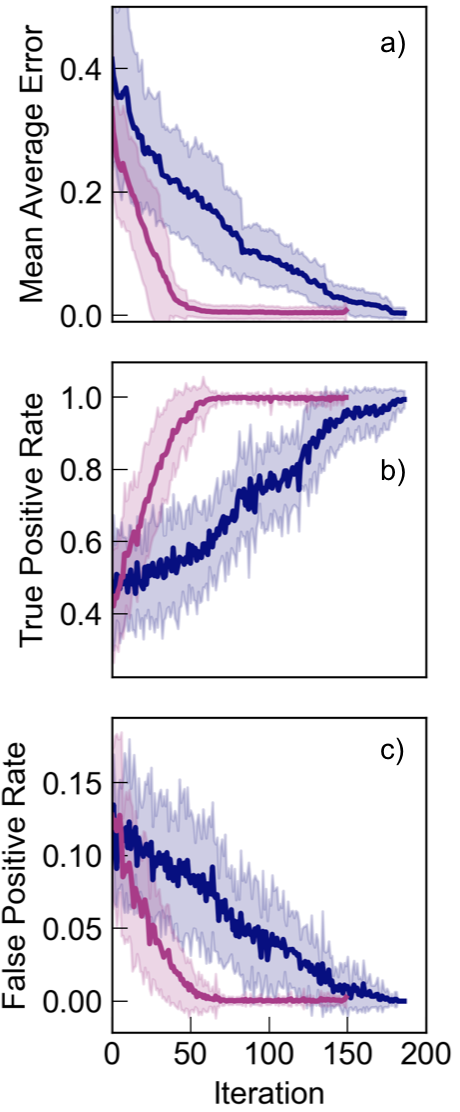}
\caption{\label{fig:Performance} 
To compare the performance of CAL (pink) and BASE (blue), we consider a more complex search problem:  ternary composition spaces with 
 three competing phases. 
(a) Concerning the regression problem for the convex hull, we calculate the average error in the convex hull energy across the composition space. (b,c) The classification accuracy is also evaluated using the true and false positive rates. Across all metrics, CAL outperforms BASE. Here, we show the performance averaged across 40 sets of energy surfaces. The bands represent one standard deviation from the mean.}
\end{center}
\end{figure} 

\begin{figure*}[t!]
\begin{center}
\includegraphics[width =0.9 \linewidth]{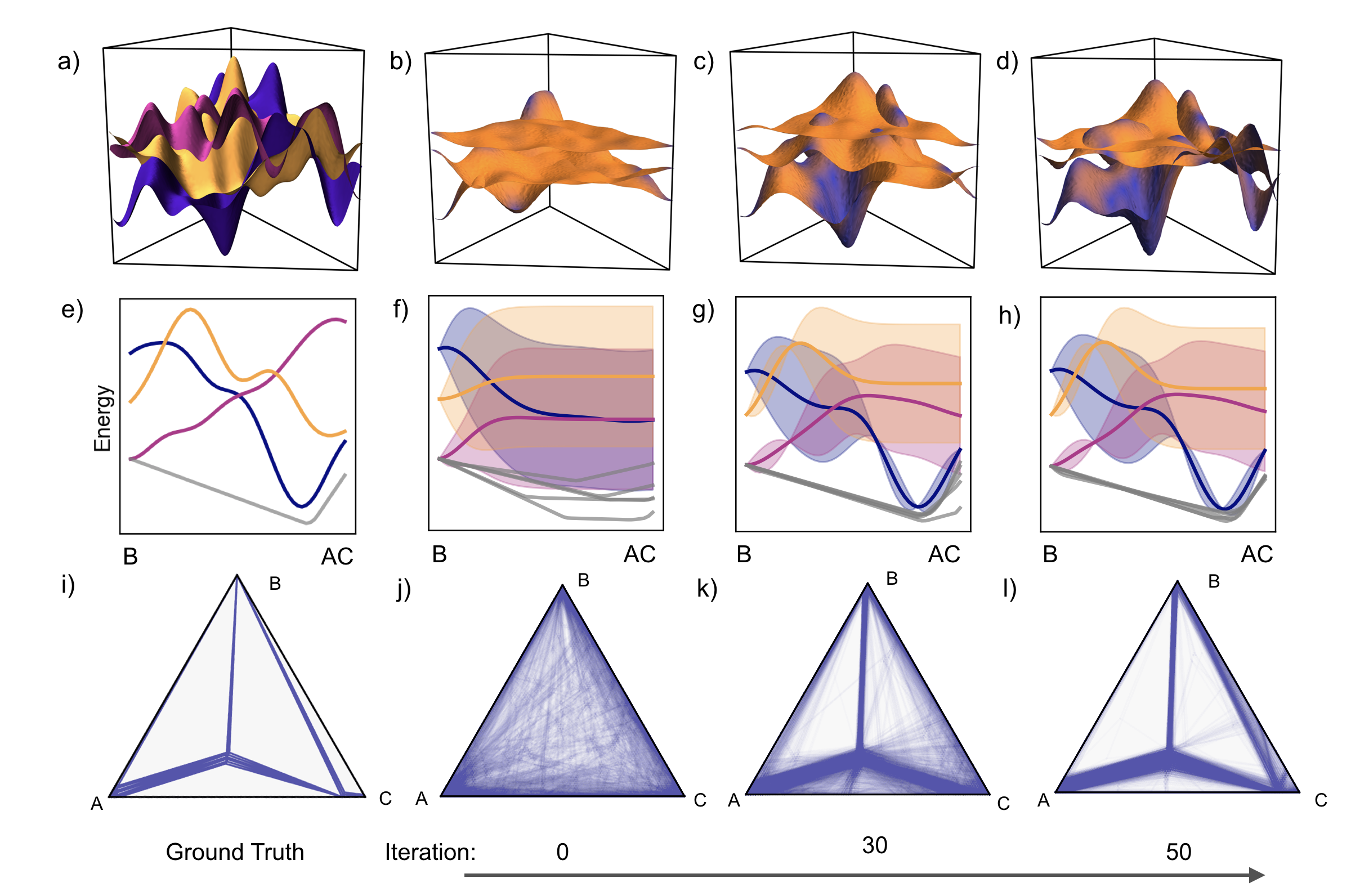}
\caption{\label{fig:2D Visualization} 
The evolution of the CAL performance is shown quantitatively in Fig~\ref{fig:Performance}; further insight can be gained by visualizing the evolution of the GP and the associated hull for a single set of energy surfaces.
To investigate how CAL performs with three phases spanning a ternary composition space (continuing Fig.~\ref{fig:Performance}), a single example is considered with increasing observations.  
(a) Each phase has an energy surface that spans the composition space. (e) A slice of the ternary space from B to AC shows the energies of these competing phases and a corresponding slice of the convex hull. (i) The full convex hull is represented as a ternary phase diagram. (b) After 10 iterations of CAL, the three Gaussian processes are illustrated by plotting their means and coloring the surfaces with their associated uncertainties. 
(c,d) With increasing iteration, CAL prioritizes learning about phase-composition pairs that are relevant to the convex hull, resulting in  regions transitioning from high (orange) to low (purple) uncertainty. (f-h) A similar progression can be seen in the slice from B to AC. 
Ultimately, we are interested in predictions of the hull and the associated phase diagram. j) After 10 iterations, the uncertainty in the convex hull distribution is represented by overlaying 100 convex hull samples on a ternary phase diagram. 
(k,l) With increasing iteration, the distribution tightens and converges around the true convex hull.}
\end{center}
\end{figure*}
\paragraph{Case Example II: Ternary Composition Space with Three Phases} 

Here we highlight a ternary composition space of the form A$_{1-x-y}$B$_x$C$_y$ with three different competing phases. This example is chosen to show how CAL navigates multiple dimensions and prioritizes phases that are more relevant to the convex hull. With composition steps of 0.1, the search space consists of 66 discrete compositions and 198 phase-composition pairs. We repeat the search process for 40 different sets of energy surfaces to reveal the typical differences between the two policies. 

Across all three metrics shown in Fig.~\ref{fig:Performance}, CAL significantly outperforms BASE. For CAL, the mean absolute error (MAE) is nearly zero by 50 iterations. Similar convergence is found for the true positive and false positive rates. Together, these metrics indicate that by 50 iterations (i.e., 25\% of the search space), CAL is able to predict the energy of the convex hull as well as classify which compositions are on and off the hull. BASE, however, takes significantly longer to come to these conclusions. Considering that there only 198 phase-composition pairs in this space, BASE requires observing nearly all phase-composition pairs to understand the convex hull. Not only does BASE finish far slower, but its rate of learning is consistently lower through the search process, as shown by its smaller slopes in Fig.~\ref{fig:Performance}a-c. Finally, from the width of the shaded regions, we conclude that BASE is much more variable than CAL.

Fig.~\ref{fig:2D Visualization} shows a representative example from Fig.~\ref{fig:Performance} to understand the root of how CAL so efficiently and consistently reveals the hull. The true energetic landscape is shown in panel (a) with energy surfaces corresponding to the three distinct phases.  
A slice through these energy surfaces is shown in (e); here, we show from B to intermediate composition AC. Additionally, a slice of the true convex hull is included below in grey. In panel (i), the complete convex hull is projected onto two dimensions as a ternary phase diagram. The three energy surfaces are similar in energy, resulting in a fairly complex phase diagram.  As such, this is a challenging task for hull determination.  

We model the three energy surfaces using separate Gaussian processes and conduct a total of 50 observations within this system.
In panels (b-d),  we show the mean of each GP and color the three surfaces by their standard deviation.  In (b), before any observations, all energy surfaces have significant uncertainty and are thus orange.  With increasing iteration, both the mean energies evolve and the uncertainties decrease for select composition regions; it will be made clear that these regions are targeted by CAL for their relevance to the convex hull. 
The evolution of energetic uncertainties can be clearly seen in the $B-AC$ slice. Composition-phase pairs near the hull show evidence of significant observation and an associated reduction in uncertainty. It is important to note that only observing the lowest energy phase would not have been an optimal solution--different phases affect the hull in different regions. 
 
In panels (j-l), 100 hulls are projected and overlaid onto the ternary phase diagram.  As expected, no coherent expectation for the hull is present initially.  
By 30 iterations, most of the single-phase regions have been identified, but there is still significant uncertainty. As such, some unstable compositions are classified as having a non-zero probability of being on the hull, resulting in a smearing out of the ternary phase diagram. Finally, after 50 iterations, much of the lingering uncertainty has dissipated and the convex hull is well understood. 

\section{Discussion}\label{sec12}
The above case examples demonstrate CAL as a fundamentally distinct approach to resolving phase diagrams. There are a variety of ways in which the general method presented can be adjusted to specific search problems. Herein, we consider joint Gaussian processes as tools for capturing correlations between separate phases. As a natural extension of joint Gaussian processes, we discuss conducting CAL simultaneously over a variety of temperatures. We then list ways in which the computational cost of CAL can be reduced for truly vast composition spaces. 

It is also explained how our method may play a role in a broader uncertainty-based thermodynamic workflow. First, the importance of uncertainty quantification is discussed, then we consider how CAL may interact with sources of uncertainty that precede it in a workflow. Finally, we talk through how the uncertainty in CAL predictions is propagated forward to other thermodynamic predictions.  

\paragraph{Correlated Energy Surfaces}
For simplicity, we used separate GPs for modeling the energy surface of each competing phase. If there are compositional correlations between energy surfaces, the set of GPs are not learning from them. For systems where strong compositional correlations are expected, it would be advantageous to use observations of one phase-composition pair to help inform the beliefs about a separate phase for similar compositions. 

Joint Gaussian processes are well-suited for incorporating compositional correlations into the energy model \cite{alvarez2012kernels,garnett2023bayesian}. In a joint Gaussian process, the energy surface of each phase would be modeled simultaneously; the inputs for such a model would be observations across all phases, and the outputs would be the energy surfaces for each phase. Incorporating joint GPs into CAL would leave the acquisition function unchanged.  

\paragraph{Temperature}
Often, it is favorable to produce phase diagrams over a range of temperatures; example applications include tuning synthesis conditions or identifying phase transitions that limit the operating conditions for a material. To incorporate temperature into the CAL workflow, the free energy surface could be modeled as a function of both composition and temperature. Such an approach would allow for the GP to explicitly learn the relationship between free energy surfaces at differing temperatures. As a terminology note, here we use the term “free energy” to explicitly denote the temperature dependence of the thermodynamic potential. 

The policy for determining the next optimal observation would need to be extended in order to account for temperature as an added dimension in the design space. The added complexity derives from the free energy convex hull only being defined over composition space at a single temperature.  As such, the total expected information gain for a single phase-composition-temperature triplet would need to be assessed as a sum over the expected information gains across temperatures of interest. In practice, the temperature range would need to be discretized to make evaluating the total information gain feasible. 

A special case of temperature-dependent search involves thermodynamic methods where calculating the enthalpy of formation is the computationally limiting factor and the entropy can be approximated analytically \cite{novick2023simulating,zunger1990special,chen2023map}. 
As such, with these methods the free energy can be predicted at multiple temperatures with no additional cost . The ramifications of this set of observations would need to be incorporated into the acquisition function. 
 
\paragraph{Computational Scaling and Approaches for Cutting Cost}

The computational cost of CAL will often be dwarfed by that of first-principles calculations. However, there is some cost to CAL, especially when moving to multi-dimensional composition spaces with many possible phase-composition pairs. If the cost of CAL is unacceptably large compared to the energy evaluations, there are multiple shortcuts for speeding up the algorithm. 

Evaluating the expected information gain (EIG) across phase-composition pairs is the main source of cost for CAL. Indeed, one could use Bayesian optimization to efficiently find the optimal phase-composition pair that maximizes the EIG. One could also imagine using a coarse grid of compositions to begin with and then iteratively increasing the granularity of the composition grid as the convex hull distribution continues to tighten. 

In truly large spaces, one may want to prioritize composition sub-regions. The acquisition function can be readily altered to exclusively focus on such regions. Here, the expected information gain would only reflect minimizing the uncertainty for the convex hull in those prioritized regions. The resulting efficiency gain will be dependent on how many different multi-phase regions enclose the specified compositions.   

Other approaches center around decreasing the cost of the EIG. For instance, the EIG could be calculated with fewer convex hull samples. Another approach would employ BASE in the beginning of the search and CAL only after some number of iterations. Since CAL is more expensive, it would be reserved for later in the search when there is sufficient information about the hull such that the CAL policy results in significantly different decisions from BASE. Finally, one could approximate the joint entropy as a sum of the entropies across individual compositions. This is a strong approximation for the entropy and should be taken with caution since it assumes convex hulls have no correlations between compositions. All these shortcuts add parameters requiring tuning to negotiate between speed and quality.

\paragraph{Opportunities for Uncertainty-based Workflows}

Understanding how uncertainty propagates throughout a workflow allows for the rational prioritization of certain segments of the workflow. Thermodynamic stability prediction is one such workflow--it often involves a series of convoluted steps, and at each step there is opportunity to estimate and propagate uncertainty. 
Such uncertainties could be produced from first-principles calculations \cite{wang2021framework,bosoni2024verify}, fitting surrogate models \cite{ober2023thermodynamically,chen2024bayesian}, or numerical approaches to approximating free energies \cite{novick2023simulating}.
The GP within CAL could incorporate uncertainties from previous steps as noise in its observations. Such noise would be reflected in the convex hull distribution and resulting predictions.  

In an uncertainty-based thermodynamic workflow, CAL could be useful in iteratively training surrogate models with energetic uncertainties like the Bayesian approach to cluster expansion \cite{mueller2009bayesian,kristensen2014bayesian,aldegunde2016quantifying,ober2023thermodynamically,chen2024bayesian}. Here, completing the necessary first-principles calculations to train such models is the limiting factor. Such training would be focused on minimizing the uncertainty in the convex hull rather than predicting energies.  

Specifically, instead of the GP used in our work, the surrogate model would be leveraged to produce uncertainty in the convex hull distribution before and after a potential observation. The simplicity of such an inexpensive surrogate makes it computationally feasible to retrain numerous times, which is necessary for choosing the optimal observation. Once an optimal composition is identified by CAL, its energy would be calculated using first-principles, and the result would be included in the training set for the surrogate model. 

\paragraph{Ultra-fine Convex Hulls}\label{Ultra-fine}
Bayesian modeling also allows for propagating uncertainty to subsequent steps in the thermodynamic workflow. We have shown such propagation for both stability predictions and chemical potentials, and herein we highlight one more example--the production of ultra-fine convex hulls from coarse-grid composition spaces. Producing fine-grained convex hulls is advantageous due to their ability to resolve single-phase regions, but conducting CAL on ultra-fine composition grids heavily increases its computational cost. As such, we use CAL to conduct search on coarse grids and use post-processing to produce the fine-grained convex hulls shown in Fig.~\ref{fig:2D Visualization}j-l. Specifically, a new GP is trained on the existing energy observations from the coarse grid and produces energetic predictions over a fine composition space. The resulting convex hull distribution is subsequently derived. The associated uncertainty with interpolating to fine grids is naturally included in the convex hull predictions. 

\section{Conclusion}\label{sec13}
Efficient, scalable calculations coupled with end-to-end uncertainty predictions are critical for the next generation of  computational materials design.  Here,  CAL provides a crucial component of this workflow with the ability  to efficiently and accurately predict thermodynamic stability. 
This enhancement comes from developing an acquisition function for active learning that is focused on minimizing the uncertainty of the convex hull. Rather than attempt to characterize the entire space, CAL prioritizes observing compositions that are on or near the hull. As a result, we see a factor of four gain in search efficiency for complex ternary spaces. While we focus on ternary spaces, our approach generalizes across dimensions; thus, it can be applied to pernicious problems such as generating phase diagrams for high-entropy alloys.  
Uncertainty quantification of both phase stability and associated intensive variables emerges naturally from this hull-aware Bayesian method.  Such intensive variables (e.g., pressure, chemical potential, voltage) are critical for linking CAL's results into a predictive workflow for informing experimental campaigns. 
 
\section{Methods}

\paragraph{Gaussian process model}

Let $\X \subseteq \reals^d$ denote the composition space; 
we assume that the composition space is a discrete set.
We model the energy surface using a Gaussian process prior:
\[
\vspace{-5pt}
F(x) \sim \mathrm{GP}(m(x), k(x,x')),
\vspace{-2pt}
\]
where $m(x)$
is the mean function and 
$k(x,x')$ is the covariance  (or kernel) function.
Given a composition $x \in \X$,  the corresponding energy is
$y~=~F(x)$.

The convex hull operator $\C$
takes an energy function $F$ and returns its lower convex envelope $H = \C(F)$.
Thus, the GP prior on the energy function $F$ implies a prior on its convex hull $H$.

Given $N$ observations $\D = \{(x_n,y_n)\}_{n=1}^N$, the
posterior of the energy function $p(F \given \D)$
 is also a Gaussian process 
$p(F \given \D) = \mathrm{GP}(\widetilde m(x), \widetilde k(x,x'))$
with mean and covariance function of the form
\begin{align}
\label{eq:gpmodel}
\widetilde m(x) &= m(x) + k_{xX} \, k_{XX}^{-1} \, (Y - m_X)
\\
\label{eq:gpmodel2}
\widetilde k(x,x') &= k(x,x') - k_{xX} \, k_{XX}^{-1} \, k_{X,x'},
\end{align}
where 
$Y = [y_1,\ldots,y_N]^\top$ is the vector of energies;   
$m_X$ and $k_{X,x'}$ are vectors of length $N$ induced from evaluating the functions $m$ and $k$ on the elements $X=\{x_n\}_{n=1}^N$; and $k_{x,X}$ and $k_{X,X}$  are matrices of dimension $1 \times N$ and $N \times N$, respectively, 
constructed from evaluating the kernel function on the elements $X$.

In practice, we represent $F$ (and $H$) using a dense grid of $c$ candidate compositions. In this case, the posterior of the energy values on this grid becomes a multivariate Gaussian distribution with a mean and covariance matrix arising from  $\widetilde m$~(\cref{eq:gpmodel}) and $\widetilde k$~(\cref{eq:gpmodel2}) evaluated at those points.

The posterior over the energy surface $F$ induces a
posterior over the convex hull function $p(H \given \D)$.
To generate a random function from this posterior, i.e.,  $H_{\D} \sim p(H \given \D)$,
we first sample 
$F_{\D}$ from $p(F \given \D)$ and then construct
 its convex hull, i.e., 
 $H_{\D} = \C(F_{\D})$.

\paragraph{Expected information gain computation}

For a given composition, CAL calculates the change in entropy for a variety of possible outcomes and averages them together to produce the expected information gain. Herein, we detail how the EIG is calculated.

Recall that to compute the EIG for the random hull $H_{\D} \sim p(H \given \D)$    (\cref{eqn:eig}),
we need to compute
the entropy
$\SS[\HD]$
and the expected entropy
$\EE_y[\SS[\HD]]$,
where 
$y$ is the (unobserved) energy of a new candidate composition $x$.
The entropy 
$\SS[\HD]$ is defined as
\begin{align}
\label{eq:entropydef}
    \SS[\HD] := -\EE_{p(H \given \D)}[\ln p(H \given \D)].
\end{align}
A key challenge is to estimate the entropy since it is  not available analytically.
In particular, in large and high-dimensional composition spaces, the expectation
in \cref{eq:entropydef} involves a 
high-dimensional integral and a  high-dimensional log hull density, both of which 
are challenging to estimate accurately and efficiently using numerical methods.

To address this computational issue,
we approximate the hull distribution in a way that allows us to compute 
\cref{eq:entropydef} in closed form.
We assume that
random values of the convex hull (evaluated on a dense grid of $c$ elements) follow a multivariate Gaussian distribution with covariance $\Sigma$.

We estimate this covariance matrix by computing the empirical covariance matrix
of $m$~convex hull samples
$ H_j 
 \sim p(H\given \D),$
 where each $H_j$ is a vector of length $c$; we  use those vectors to construct
the covariance matrix: 
\begin{equation}
    \Sigma=\tfrac{1}{m}\textstyle\sum_{j=1}^m (H_j-\bar{H}) \, (H_j - \bar{H})^\top.
\end{equation}
Here, $\bar{H} := \tfrac{1}{m}\sum_{j=1}^m H_j$
is the vector obtained from averaging over the components of each hull vector $H_j$.
It is worth noting that the number of convex hull samples must satisfy $m>c$ in order to ensure the covariance matrix is full-rank.

The entropy of the multivariate Gaussian,
which only depends on the covariance $\Sigma$, can be computed in closed form:
\begin{align}
\label{eq:entropy}
\SS[\HD] =
\tfrac{c}{2}\ln(2\pi e) + \tfrac{1}{2} \ln(\det(\Sigma)).
\end{align}

For the expected entropy $\EE_y[\SS[\HD]]$,
we compute a Monte Carlo estimate
of the expectation: 
\begin{align}
    \EE_y[\SS[\HDnew]]
    \approx
    \tfrac{1}{K} \textstyle\sum_{k=1}^K
    \SS[\HDnewk],
\end{align}
where $\{y_k\}_{k=1}^K$ are samples 
obtained from  
the posterior $p(F\given \D)$ for a given composition $x$,
and the entropy estimates
$\SS[\HDnewk]$ are computed using \cref{eq:entropy}.

\paragraph{Implementation and evaluation details}
The Gaussian process model and active search algorithm were implemented using \texttt{JAX} \cite{JAX} and the \texttt{GPJax} library \cite{GPJAX}. For simplicity and consistency, a radial basis function (RBF) kernel with a length scale of 0.2 was used throughout the paper. This length scale was chosen as it gave energy curves that generally agreed with other thermodynamic potentials. The ``true" energy surfaces were generated using an RBF kernel with a length scale of 0.2 as well. 

All observations had no noise associated with them, although observational noise can readily be incorporated. Shaded regions in the GP plots show two standard deviations from the mean prediction. Convex hulls were generated using the \texttt{qhull} algorithm \cite{qhull} within the scipy library \cite{scipy}. Custom code was built to isolate the lower bound of the hull, which is the portion of interest for thermodynamics.

Here we discuss the specific sampling parameters used in the work. In the 1D search evolution shown in Fig.~\ref{fig:1D Time Series}, there were 21 compositions in the space. For the ternary search in Fig.~\ref{fig:Performance} and Fig.~\ref{fig:2D Visualization}, there were 66 total compositions. For both the 1D and 2D search, 200 energy and convex hull samples were used for each entropy calculation, and 10 possible $y$-values were used to build the expected information gain (i.e., $m=200$, $K=10$).

The baseline active learning algorithm, which also uses a GP model for the energy surface, selected compositions to maximize the information gained about the energy surface. 
Specifically, BASE maximized
the EIG with respect to the energy function (EIG-B):
\begin{align*}
\label{eqn:eigbase}
    \text{EIG-B}(x\,;\,\mathcal{D}) &:=
    \SS[F_{\mathcal{D}}]
    -\mathbb{E}_{y}[\SS[F_{\mathcal{D}\cup(x,y)}]]\,.
\end{align*}

\noindent When multiple phases were present, BASE chose the composition-phase pair that maximized the EIG. In Fig.~\ref{fig:Performance}, the policy resulted in BASE alternating evenly between phases. BASE used the same GP hyperparameters as CAL to control for hyperparameter tuning. 

The performance of each policy was assessed using the mean absolute error (MAE) of the hull energy, the true positive rate (TPR), and false positive rate (FPR). The MAE here is defined by:
\begin{equation}
    MAE = \tfrac{1}{c}\textstyle\sum_{i}^c |\mathbb{E}_m[H_{i,pred}]-H_{i,true}|.
\end{equation}
For composition $i$, the error is defined as the absolute difference between the average predicted hull energy ($H_{i,pred}$) and the true hull energy ($H_{i,true}$). The absolute value of these errors is then averaged over all compositions, $c$. 

The true positive rate is the percentage of the points that are on the hull that are correctly identified: 
\begin{equation}
    TPR = \frac{TP}{TP+FN}.
\end{equation}
TP refers to the number of true positives, which in this context is the number of compositions that are correctly identified as being on the convex hull. FN is the number of false negatives, which is the number of compositions that were incorrectly identified as being off the hull. 

The FPR refers to the percentage of the points that are off the hull that were incorrectly identified: \begin{equation}
    FPR = \frac{FP}{FP+TN}
\end{equation}   
FP is the number of false positives, which is the number of compositions that were incorrectly identified as being on the convex hull. TN stands for true negative, and is the number of compositions that were correctly identified as being off the hull. 

For both CAL and BASE, 200 hulls were used to evaluate the MAE, TPR, and FPR for a given iteration. A composition was defined as being on the hull if its energy was within $10^{-3}$ of the energy of the hull. 

\section{Acknowledgments}
This work was supported under NSF OAC 2118201.  Additional support came from NSF OAC 1940199 (AN and ET), NSF IIS 1845434, NSF IIS 2007278 and NSF OAC 1940224 (QN and RG), NSF IIS 2007278 (RA), and a Google Ph.D.\ Fellowship in Machine Learning (DC).  
\vspace{3cm}

\backmatter

\printbibliography

\end{document}